\documentclass[twocolumn,groupedaddress,amsmath,amssymb,prb,aps]{revtex4}

\usepackage{cabin}
\usepackage{graphicx}
\usepackage{dcolumn}
\usepackage{bm}
\usepackage{soul} 
\usepackage{xcolor} 
\usepackage{amsmath} 

\begin{document}

\title{Giant voltage amplification from incipient ferroelectric states}

\author{M\'onica Graf$^{1}$, Hugo Aramberri$^{1}$, Pavlo Zubko$^{2}$
  and Jorge \'I\~niguez$^{1,3}$}

\affiliation{
  \mbox{$^{1}$Materials Research and Technology Department,
    Luxembourg Institute of Science and Technology (LIST),} \mbox{Avenue
    des Hauts-Fourneaux 5, L-4362 Esch/Alzette,
    Luxembourg}\\
\mbox{$^{2}$London Centre for Nanotechnology and Department of Physics and
  Astronomy,} \mbox{University College London, 17–19 Gordon Street, WC1H 0HA
  London, United Kingdom}\\
 \mbox{$^{3}$Department of Physics and Materials Science, University
   of Luxembourg, Rue du Brill 41, L-4422 Belvaux, Luxembourg}}

\maketitle   

\textbf{Ferroelectrics subject to suitable electric boundary
  conditions present a steady negative capacitance response. When the
  ferroelectric is in a heterostructure, this behavior yields a
  voltage amplification in the other elements, which experience a
  potential difference larger than the one applied, holding promise
  for low-power electronics. So far research has focused on verifying
  this effect and little is known about how to optimize it. Here we
  describe an electrostatic theory of ferroelectric/dielectric
  superlattices, convenient model systems, and show the relationship
  between the negative permittivity of the ferroelectric layers and
  the voltage amplification in the dielectric ones. Then, we run
  simulations of PbTiO$_{3}$/SrTiO$_{3}$ superlattices to reveal the
  factors most strongly affecting the amplification. In particular, we
  find that giant effects (up to 10-fold increases) can be obtained
  when PbTiO$_{3}$ is brought close to the so-called ``incipient
  ferroelectric'' state.}



Since its revival in 2008\cite{salahuddin08}, the negative capacitance
(NC) response of ferroelectrics has been a focus of attention. In
principle, all materials must present a positive global capacitance or
dielectric constant, a necessary condition for thermodynamic
stability. Nevertheless, local NC states can be obtained in a variety
of ways\cite{iniguez19}. Most interestingly, by placing a
ferroelectric in contact with a dielectric, we can prevent it from
reaching its lowest-energy state (homogenous polarization), forcing it
to display a different order of relatively high energy. Such a
frustrated ferroelectric will typically display a steady NC response
upon application of an electric
field\cite{bratkovsky01,bratkovsky06,iniguez19}. This has been shown
in detail for ferroelectric/dielectric superlattices that have emerged
as model systems in the field\cite{zubko16,yadav19,das21,pavlenko22}.


\begin{figure}
\includegraphics[width=0.5\columnwidth]{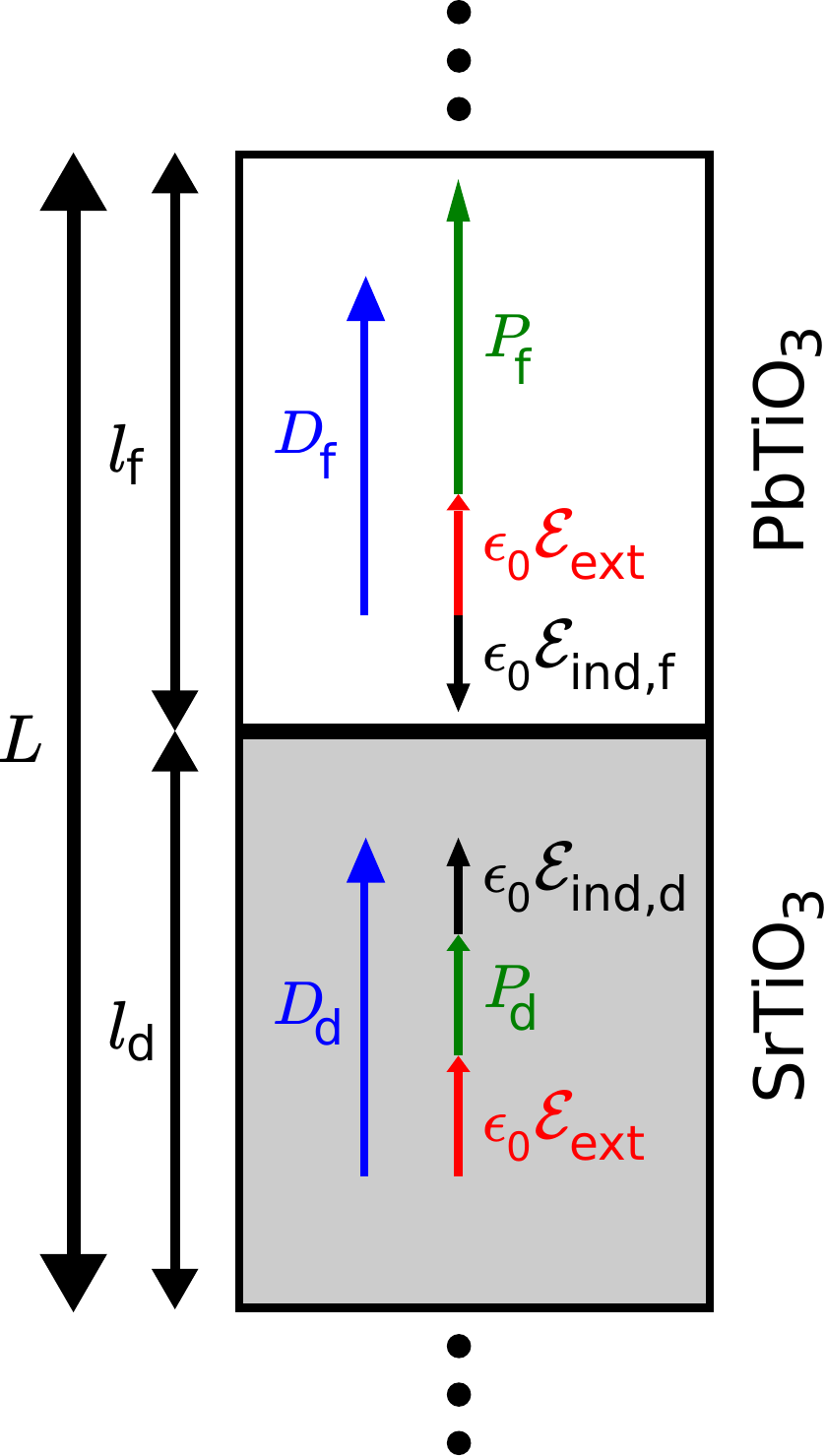}
\caption{Sketch of a ferroelectric/paraelectric superlattice
  periodically repeated along the stacking direction. The thickness of
  the ferroelectric and dielectric layer is given by $l_{\rm f}$ and
  $l_{\rm d}$, respectively; $L = l_{\rm f}+l_{\rm d}$ is the
  thickness of the repeated unit. For an arbitrary external field
  $\mathcal{E}_{\rm ext}$, and in the absence of free carriers, all layers
  present the same vertical component of the displacement vector, so
  that $D_{\rm f} = D_{\rm d}$. As illustrated in the figure, the
  displacement $D_{i}$ of layer $i$ involves the layer polarization
  $P_{i}$, the field $\mathcal{E}_{{\rm ind},i}$ induced in the layer
  and the external field $\mathcal{E}_{\rm ext}$.}
\label{fig:scheme}
\end{figure}

To understand steady-state NC, let us consider the superlattice in
Fig.~\ref{fig:scheme}, where ferroelectric (f) and dielectric (d)
layers repeat periodically along the stacking direction $z$. In the
absence of free carriers, the continuity of the $z$ component of the
displacement vector implies $D = D_{\rm f} = D_{\rm d}$, where $D$ is
the superlattice displacement while $D_{\rm f}$ and $D_{\rm d}$ are
the layer vectors (we omit the $z$ subscript for simplicity). Using
the definitions in Fig.~\ref{fig:scheme}, this yields
\begin{equation}
P + \epsilon_0 \mathcal{E}_{\rm ext} = P_{\rm f}+\epsilon_0
\mathcal{E}_{\rm f} = P_{\rm d}+\epsilon_0 \mathcal{E}_{\rm d} \; ,
\end{equation}
where $\epsilon_{0}$ is vacuum permittivity, $P = L^{-1}(l_{\rm
  f}P_{\rm f}+l_{\rm d}P_{\rm d})$ is the superlattice polarization,
$\mathcal{E}_{\rm ext}$ is the external electric field along $z$, and
the total field in layer $i$ ($i =$~f, d) is 
\begin{equation} 
  \mathcal{E}_{i} = \mathcal{E}_{\rm ext} + \mathcal{E}_{{\rm ind},i}
  \; .
\end{equation}
Further, since $D=D_{i}$ we can write
\begin{equation}
    \mathcal{E}_{{\rm ind},i} = \epsilon_0^{-1}\left(P-P_{i}\right) \; ,
    \label{eq:Eind}
\end{equation}
which shows that induced fields $\mathcal{E}_{{\rm ind},i}$ appear
when the local and global polarizations differ. For the f-layer we
typically have $P_{\rm f}>P$, so that $\mathcal{E}_{\rm ind,f}$
opposes $P_{\rm f}$; this is the so-called ``depolarizing field''.

Because of the superlattice periodicity, the total voltage associated
to the induced fields must be null, which implies $l_{\rm
  d}\mathcal{E}_{\rm ind,d}+l_{\rm f}\mathcal{E}_{\rm ind,f}=0$. This
means that there is no net depolarizing field, $\mathcal{E}_{\rm ext}$
being the only macroscopic field acting on the system.

To examine the response to a variation of the external field
$d\mathcal{E}_{\rm ext}$, it is useful to introduce a quantity we call
the ``screening factor'', defined for the f-layer as
\begin{equation}
    \varphi_{\rm f} = \frac{d \mathcal{E}_{\rm ind,f}}{d
      \mathcal{E}_{\rm{ext}}} =
    \epsilon_{0}^{-1}\frac{d\left(P-P_{\rm f}\right)}{d\mathcal{E}_{\rm{ext}}}
    = \frac{l_{\rm d}}{L}\left(\chi'_{\rm d}-\chi'_{\rm f}\right)\; .
    \label{eq:varphi}
\end{equation}
Here we use the primed susceptibilities $\epsilon_{0}\chi'_{i} =
dP_{i}/d\mathcal{E}_{\rm ext}$, which are all but guaranteed to be
positive. (The change in polarization -- local or global -- will
always follow the change in the external field. By contrast, the usual
susceptibility of the f-layer $\epsilon_{0}\chi_{\rm f} = dP_{\rm
  f}/d\mathcal{E}_{\rm f}$ involves the local field, and becomes
negative in the NC regime.) The inverse permitivitty of the f-layer
can then be written as
\begin{equation}
    \epsilon_{\rm f}^{-1} = \frac{d\mathcal{E}_{\rm f}}{dD} =
    \frac{d\mathcal{E_{\rm ext}}}{dD} \left(1+\varphi_{\rm f}\right) = 
    \epsilon^{-1}\left(1+\varphi_{\rm f}\right) \; .
    \label{eq:epsilon_f} 
\end{equation}
Further, as detailed in Supplementary Note~1, we can derive the
voltage response of the dielectric layer $\mathcal{A}_{\rm d}$ as
\begin{equation}
  \mathcal{A}_{\rm d} = \frac{dV_{\rm d}}{dV} =
  \frac{l_{\rm d}}{L}\frac{d\mathcal{E}_{\rm d}}{d\mathcal{E}_{\rm ext}} =
  L^{-1}\left(l_{\rm d}-l_{\rm f}\varphi_{\rm f}\right) \; .
  \label{eq:A}
\end{equation}
Voltage amplification (VA) corresponds to $\mathcal{A}_{\rm
  d}>1$. This key quantity is fully determined by trivial geometric
elements and the screening factor of the f-layer.

We now use these equations to discuss the dielectric response of a
superlattice. Typically the ferroelectric layers will be more
responsive than the dielectric ones, so that $\chi'_{\rm f} >
\chi'_{\rm d}$. From Eq.~(\ref{eq:Eind}) we get that the induced
depolarizing field $d\mathcal{E}_{\rm ind,f}$ will oppose
$d\mathcal{E}_{\rm ext}$, and hence $\varphi_{\rm f}<0$. One usually
expects the induced field to be smaller in magnitude than the applied
one, so that $-1<\varphi_{\rm f}<0$. It follows that $\epsilon_{\rm
  f}^{-1}>0$ and $\mathcal{A}_{\rm d}<1$, a behavior we may call
normal.

Imagine we make the ferroelectric more responsive, e.g., by varying
its temperature to approach a Curie point. Then, we can eventually
reach a situation where the induced $d\mathcal{E}_{\rm ind,f}$
compensates the applied $d\mathcal{E}_{\rm ext}$ ($\varphi_{\rm
  f}=-1$), and the voltage drops exclusively in the dielectric layers
($\mathcal{A}_{\rm d}=1$). The ferroelectric effectively behaves as a
metal, and we call this ``perfect screening''.

If we keep softening the f-layer so that $\chi'_{\rm f}\gg\chi'_{\rm
  d}$, we access a regime where the ferroelectric ``over-screens'':
its response is so strong that the induced depolarizing field exceeds
the applied one ($\varphi_{\rm f}<-1$). This yields NC ($\epsilon_{\rm
  f}^{-1}<0$) and VA in the dielectric ($\mathcal{A}_{\rm d}>1$).


The above formulas show that NC and VA can be obtained from the layer
polarizations, easily accessible from atomistic simulations. The
so-called ``second-principles'' methods\cite{wojdel13,scaleup} (see
Methods) have been key to explain NC in PbTiO$_{3}$/SrTiO$_{3}$
(PTO/STO) superlattices. Here we use said methods to monitor the
dependence of NC and VA on the design variables offered by these
artificial materials (layer thickness, epitaxial strain), unveiling a
distinct strategy to obtain giant effects.

We study PTO/STO superlattices where the PTO and STO layers have a
thickness of $n$ and $m$ perovskite cells, respectively, denoted
$n$/$m$ in the following. We consider $n$ and $m$ from 3 to 9, and
investigate the response to small external fields along $z$. We also
vary the epitaxial strain $\eta$ between $-1$~\% to $+3$~\%, choosing
the STO substrate as the zero of strain.

For computational feasibility we restrict ourselves to low
temperatures (formally, 0~K) and work with periodically-repeated
supercells that are relatively small in plane ($8\times 8$ perovskite
units). As argued below, this suffices to draw conclusions on the
behavior of real materials at ambient conditions.

\begin{figure}
    \centering \includegraphics[width=\columnwidth]{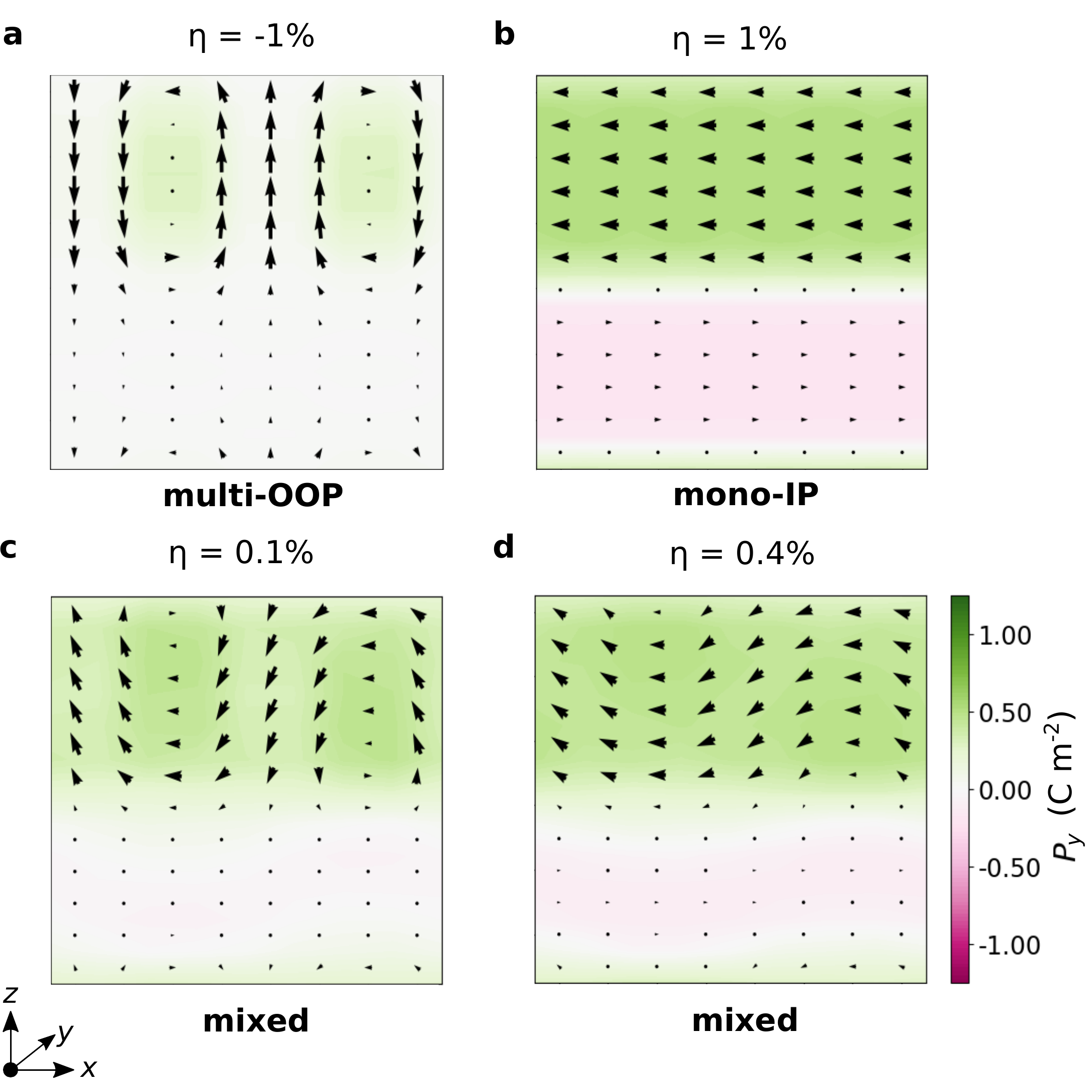}
    \caption{Representative states found for 6/6
      superlattices. Multi-OOP state for $\eta = -$1~\% shown in panel
      {\bf a}, mono-IP state for $\eta =$1~\% in panel {\bf b}, and
      mixed states for $\eta =$ 0.1~\% and 0.4~\% in panels {\bf c}
      and {\bf d}, respectively. Arrows represent local polarization
      in the $xz$ plane and the color scale corresponds to the
      polarization along $y$.}
    \label{fig:states}
\end{figure}

Let us first recall the main effect epitaxial strain has on PTO/STO
superlattices, as obtained from our
simulations. Figure~\ref{fig:states}{\bf a} shows the ground state of the
6/6 system for $\eta = -1$~\%: it presents stripe domains in the PTO
layer, with local polarizations along the out of plane (OOP) $z$
direction. This ``multi-OOP'' vortex-like state has been thoroughly
studied in the literature
\cite{zubko10,aguadopuente12,yadav16,zubko16,das19,baker20}.

For large enough tensile strains, we find the PTO layer displays a
monodomain state with in-plane (IP) polarization
(Fig.~\ref{fig:states}{\bf b}). This simulated ``mono-IP'' configuration is
characterized by $P_{x}=P_{y}$. In reality\cite{damodaran17} one
typically observes the so-called $a_{1}/a_{2}$ multidomain
configuration, with local polarizations alternating between $P_{x}$
and $P_{y}$. Our monodomain result is a consequence of the relatively
small size of the simulation supercell.

Finally, Figs.~\ref{fig:states}{\bf c} and \ref{fig:states}{\bf d} show states we
find in some superlattices at intermediate $\eta$ values, where
mono-IP and multi-OOP features ``mix'', reminiscent of similar
findings in the literature\cite{damodaran17,baker20}. In conclusion,
apart from some non-essential size effects, our simulations capture
the evolution of PTO/STO superlattices with epitaxial strain.

\begin{figure}
\includegraphics[width=\columnwidth]{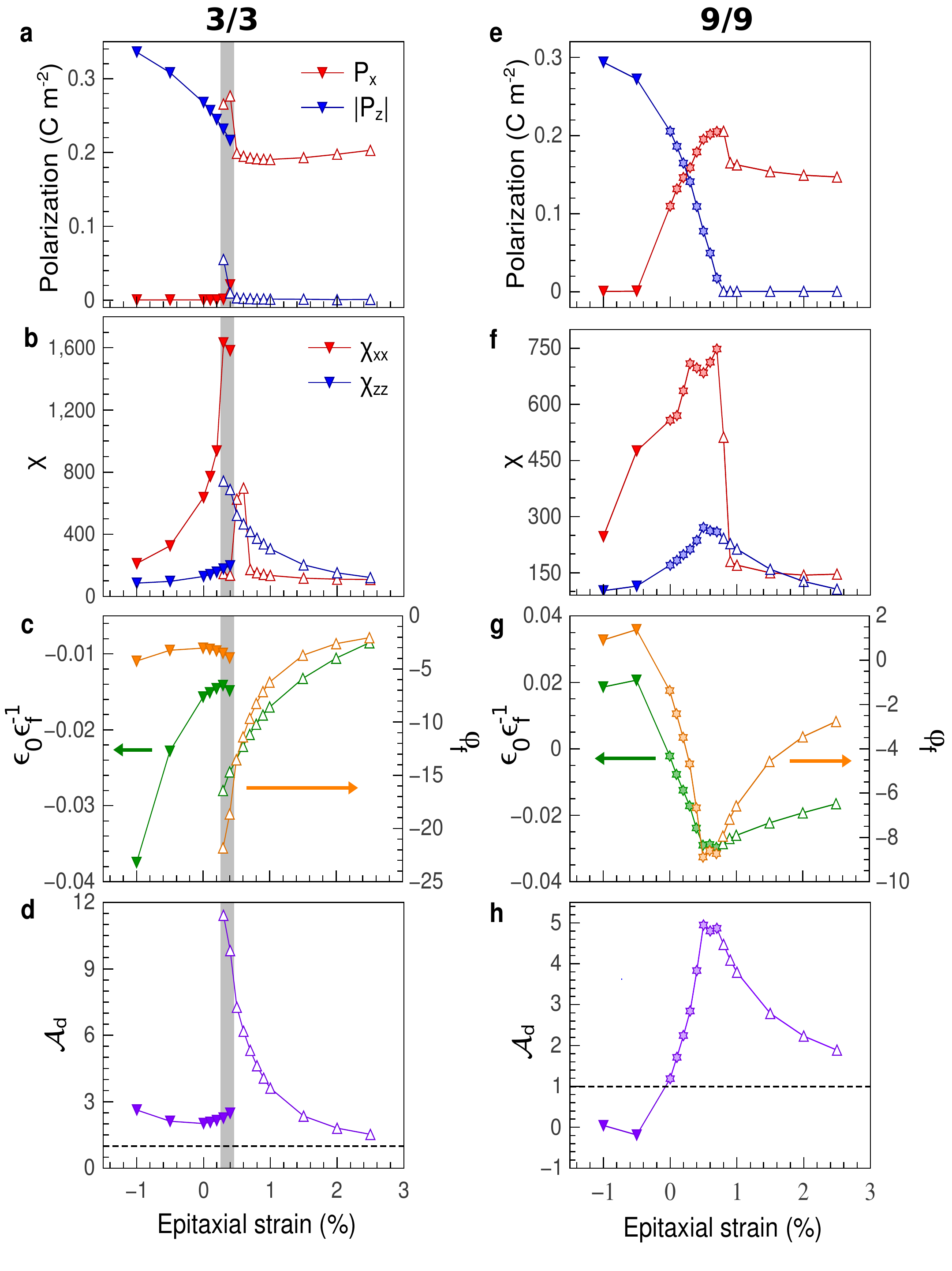}
\caption{Simulation results for the 3/3 (from {\bf a} to {\bf d}) and
  9/9 (from {\bf e} to {\bf h}) superlattices, as a function of
  epitaxial strain. Panels~{\bf a} and {\bf e} show two superlattice
  averages of the polarization: $\left|P_{z}\right|$ corresponds to
  averaging the absolute value of the $z$-component of the local
  polarizations, so as to get a non-zero result in the multi-OOP
  state; $P_{x}$ is the direct supercell average of the $x$-component
  of the local polarizations, where $x$ is the modulation direction
  (perpendicular to the domain walls) in the multi-OOP and mixed
  states. Panels~{\bf c} and {\bf g} show the inverse permittivity
  $\epsilon_{\rm f}^{-1}$ in units of $\epsilon_{0}^{-1}$ (left axis)
  and screening factor $\varphi_{\rm f}$ (right axis) of the
  ferroelectric layer. Panels~{\bf d} and {\bf h} show the voltage
  ratio $\mathcal{A}_{\rm d}$ of the dielectric. The gray zone in
  panels {\bf a}-{\bf d} marks the region where both multi-OOP and
  mono-IP states are (meta)stable. Dark-colored down-pointing
  triangles correspond to multi-OOP states, while we use light-colored
  stars for mixed states and empty up-looking triangles for mono-IP
  states.}
\label{fig:results}
\end{figure}

Figures~\ref{fig:results}{\bf a}-{\bf d} show detailed results for the
3/3 system. At compressive and slightly tensile strains, we get a
multi-OOP solution similar to that of Fig.~\ref{fig:states}{\bf a},
with $|P_{z}|\neq0$ and $P_{x}=0$. As $\eta$ increases, we see an
abrupt transition to the mono-IP phase with $|P_{z}|=0$ and $P_{x}\neq
0$. As typical of discontinuous transformations, the multi-OOP and
mono-IP states are both stable for intermediate strains (gray area in
the figure).

The global dielectric susceptibility is shown in
Fig.~\ref{fig:results}{\bf b}. As we increase $\eta$ in the multi-OOP
state, we induce a maximum of $\chi_{xx}$, signaling the occurrence of
an IP polar instability. In the case of the mono-IP state, it is
$\chi_{zz}$ that peaks as $\eta$ decreases, indicating a soft OOP
polar mode. The mono-IP state also displays a peak in $\chi_{xx}$ at
$\eta\approx 0.6$~\%; this feature, associated to the STO layer and
not essential here, is discussed in Supplementary Note~2 and
Supplementary Figure~1.

Figure~\ref{fig:results}{\bf c} shows the inverse permittivity (green)
and screening factor (orange) of the f-layer. For all considered
strains we get $\epsilon_{\rm f}^{-1}<0$ and the associated
over-screening ($\varphi_{\rm f} < -1$). Further,
Fig.~\ref{fig:results}{\bf d} shows the corresponding VA in the
d-layer, with $\mathcal{A}_{\rm d}$ reaching values as high as 12 when
the mono-IP state approaches its stability limit. This giant
amplification is related to the maximum in $\chi_{zz}$ (panel~{\bf
  b}), in turn connected to the OOP polar instability of the PTO
layer. In contrast, the destabilization of the multi-OOP state upon
increasing $\eta$ -- which involves a $\chi_{xx}$ anomaly -- does not
result in any feature in $\epsilon_{\rm f}^{-1}$ or $\mathcal{A}_{\rm
  d}$.

As shown in Figs.~\ref{fig:results}{\bf e}-{\bf h}, the 9/9
superlattice presents a similar behavior, the main difference being
that we find no $\eta$ values where two states can exist. Instead, for
$\eta$ between 0.0~\% and 0.8~\% we see a gradual transformation from
multi-OOP to mono-IP, with the occurrence of the mixed state mentioned
above (Figs.~\ref{fig:states}{\bf c} and \ref{fig:states}{\bf d}). The
small jump in $P_{x}$ around $\eta = 0.9$~\% is related to the
occurrence of an IP polarization in the STO layer (not relevant here;
see Supplementary Note~2 and Supplementary Figure~2).

The 9/9 superlattice displays its largest NC response in this
intermediate region, reaching 5-fold amplifications at the transition
point between the mono-IP and mixed states. The VA extends into the
mono-IP region.

In contrast, the multi-OOP state of the 9/9 superlattices shows a
different behavior: see e.g. $\mathcal{A}_{\rm d}<0$ at $\eta =
-0.5$~\% in Fig.~\ref{fig:results}{\bf h}. In this regime, the PTO layer is
in a very stable (stiff) multidomain configuration, while the
in-plane compression makes STO electrically soft along $z$. Hence, the
roles reverse and the STO layer displays NC. (See Supplementary Note~3
for more.) A similar behavior has been predicted in
BaTiO$_{3}$/SrTiO$_{3}$ superlattices\cite{walter20}.

\begin{figure}
    \centering
    \includegraphics[width=\columnwidth]{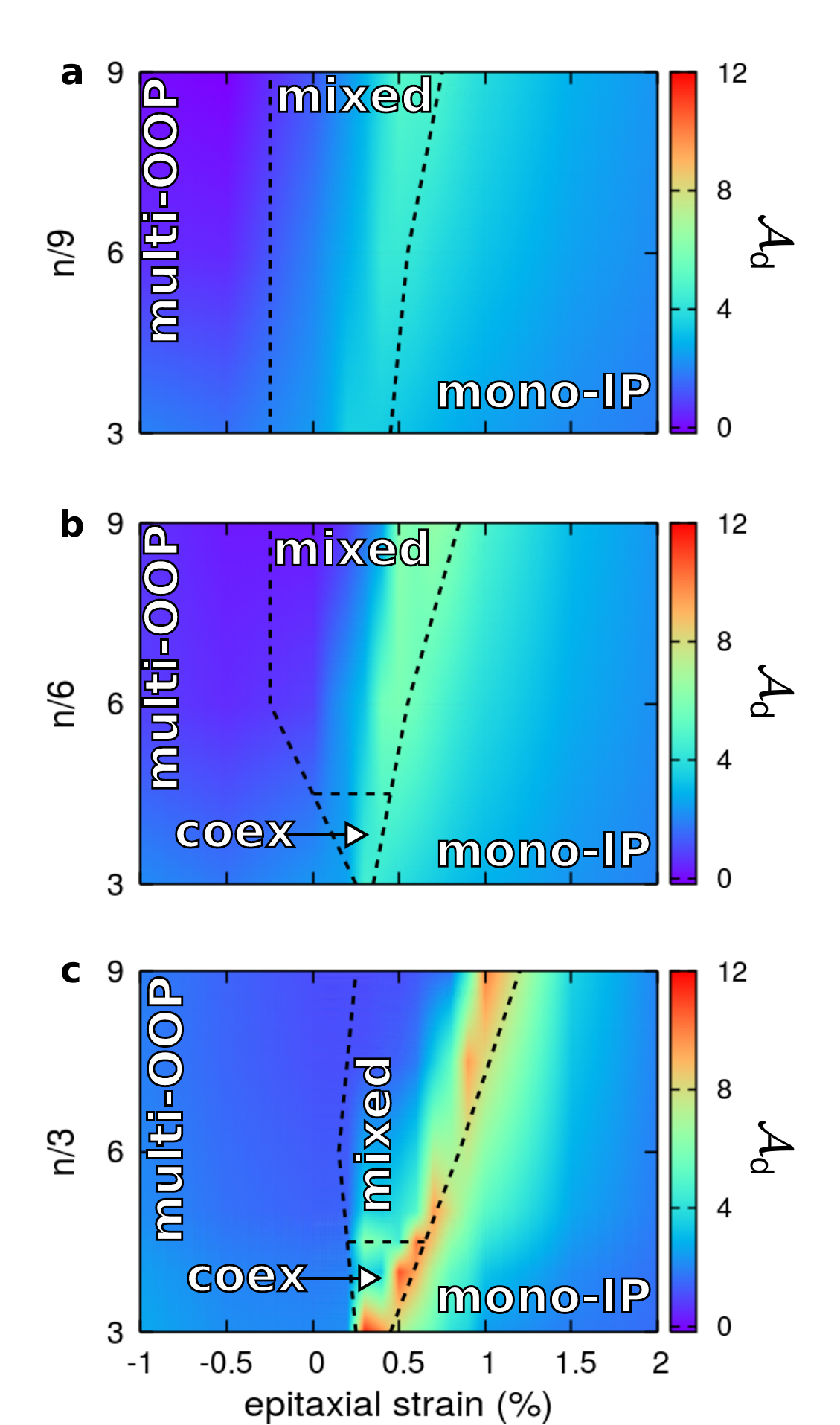}
    \caption{Summary of our results for VA in $n$/$m$
      superlattices. Panels {\bf a}, {\bf b} and {\bf c} present
      results for $m = 3$, $m = 6$ and $m = 9$, respectively. The
      color scale represents the voltage ratio $\mathcal{A}_{\rm d}$.
      The lines and labels indicate the stability regions of the
      states of Fig.~\ref{fig:states}. In the coexistence region we
      show the $\mathcal{A}_{\rm d}$ values corresponding to the
      mono-IP state.}
    \label{fig:phase_diagram}
\end{figure}

We run the same study for a large collection of superlattices;
Figure~\ref{fig:phase_diagram} summarizes our results.

We find that the transition region between the multi-OOP and mono-IP
states becomes wider for thicker PTO. This reflects the fact that
broader PTO layers can accommodate more complex dipole orders, as the
one occurring in the mixed state. (This is consistent with recent
results in the literature, e.g. the occurrence of supercrystals in
PbTiO$_{3}$/SrRuO$_{3}$ superlattices with PTO layers above 15
cells\cite{hadjimichael21}.)

The mixed state is also favored by thicker STO layers. This effect is
more subtle, and probably related to the fact that the stray fields
are expelled from the STO layer as it thickens. This aspect is not
essential here and we do not pursue it.

Most importantly, Fig.~\ref{fig:phase_diagram} confirms that the
strongest amplifications occur at the stability limit of the mono-IP
state. It also shows that the multi-OOP region is comparatively
unresponsive. Let us now get insight into the physical underpinnings
of these behaviors.


As captured by Eqs.~\ref{eq:varphi} and \ref{eq:A}, VA is essentially
determined by the screening factor of the f-layer, which in turn
depends on the difference in dielectric response between layers. For
example, for the 3/3 superlattice at $\eta = 0.3$~\% we get
$\mathcal{A}_{\rm d}\approx 12$, with $\chi'_{\rm f} = 765$ and
$\chi'_{\rm d} = 721$. This $\chi'_{\rm f}$ value may seem small;
indeed, the ferroelectric is close to developing an OOP polar
instability and, in such conditions, one expects susceptibilities
$\chi_{\rm f} > 10,000$\cite{lines-book1977,graf21}. By contrast, the
computed $\chi'_{\rm d}$ is surprisingly large, as our model for STO
yields $\chi = 202$ for the pure material. (Our simulated STO is
relatively stiff as compared to experiment\cite{zubko16}; not
essential here.)

The reason for these surprising $\chi'_{i}$ susceptibilities can be
traced back to the electrostatic requirement that $D_{\rm f}=D_{\rm
  d}$. This leads all layers to respond similarly to an external
field, to minimize the depolarizing fields. Thus, we expect
$\chi'_{\rm f}\gtrsim \chi'_{\rm d}$. For example, for the 6/6
superlattice at $\eta = -1$~\%, which does not display VA, we obtain
$\chi'_{\rm f}=96$ and $\chi'_{\rm d}=95$ (Supplementary
Figure~3). Then, when we move to a region of the phase diagram where
the f-layer presents an OOP instability, the energy gain associated to
the development of $dP_{\rm f}$ overwhelms the cost of creating a
depolarizing field. Hence, the difference between $\chi'_{\rm f}$ and
$\chi'_{\rm d}$ grows a little, which suffices to yield large VA
values.

The largest amplifications correspond to the region marking the limit
of stability of the mono-IP state. We can say that, in this area, the
f-layers are in an ``incipient ferroelectric''
state\cite{iniguez19,walter20}: they are unstable against the
development of an homogeneous OOP polarization whose occurrence is
precluded by the presence of the d-layers. Eventually, as we move
towards negative $\eta$ values, the multi-OOP polar instability
freezes in, leading to either a pure multi-OOP state or a mixed state,
and progressively hardening the $z$-polarized ferroelectric soft
mode. (This resembles the competition between antipolar and polar
orders in antiferroelectrics\cite{kittel51,lu20}.)  This incipient
ferroelectric state corresponds to the idealized picture of monodomain
NC\cite{salahuddin08,iniguez19}; our results predict a realization of
this archetype.

\begin{figure}
    \centering
    \includegraphics[width=\columnwidth]{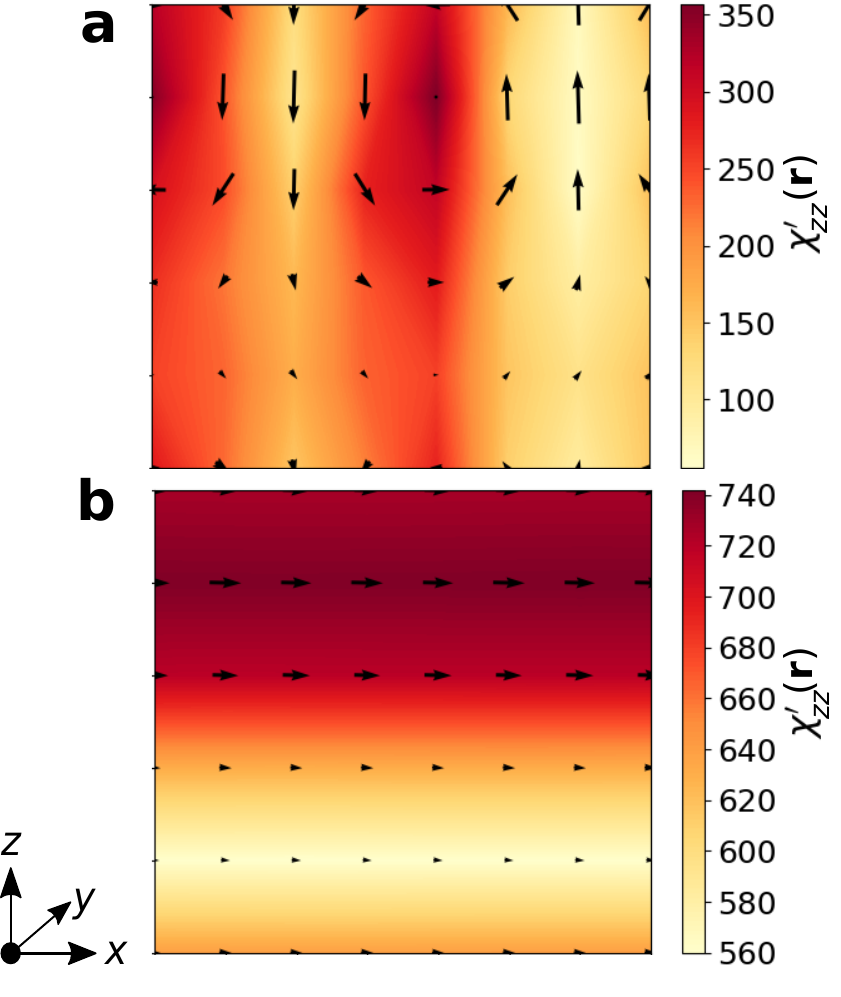}
    \caption{Maps of the local dielectic response $\chi'_{zz}({\bf
        r})=\epsilon_{0}^{-1}\partial P({\bf
        r})/\partial\mathcal{E}_{\rm ext}$, where $P({\bf r})$ is the
      position dependent $z$-component of the polarization and the
      applied field is also along $z$. The results correspond to a
      particular $xz$ plane of the 3/3 superlattice at $\eta =
      0.4$~\%. (These structures are periodic along $y$.) The arrows
      represent the local electric dipoles in the $xz$ plane at zero
      field. The shown multi-OOP ({\bf a}) and mono-IP ({\bf b})
      states are both stable for this value of $\eta$. Note that the
      color scales differ between panels.}
    \label{fig:local-chi}
\end{figure}

As shown in Fig.~\ref{fig:local-chi}{\bf a} and already
reported\cite{zubko16,yadav19}, the NC response of multi-OOP states
mainly stems from the large dielectric response (large $\chi'$) of the
domain walls. By contrast, in the incipient ferroelectric state the NC
response comes from the full volume of the f-layer
(Fig.~\ref{fig:local-chi}{\bf b}). This partly explains the superior
VA performance of the mono-IP state.

Our results suggest a strategy to obtain large VA: work with incipient
ferroelectric states that will typically occur at the boundary between
IP and OOP phases. Phase boundaries akin to the ones discussed here have
been found experimentally in PTO/STO superlattices\cite{damodaran17}
and predicted in other ferroelectric/dielectric
heterostructures\cite{walter20}; those are clear candidates to display
giant VA effects. Note that, despite their limitations (low
temperature, small supercells), our simulations capture the main
physics of the IP-to-OOP transtion; thus, our conclusions can be
expected to apply to experimentally relevant situations.

Additionally, our electrostatic formulas teach us that
$\mathcal{A}_{\rm d}$ does not depend on the macroscopic permittivity
$\epsilon^{-1}$ (Eq.~\ref{eq:A}), while $\epsilon_{\rm f}^{-1}$ does
(Eq.~\ref{eq:epsilon_f}). As a consequence, one can have behaviors as
that of the 3/3 system at $\eta = -1$~\% (Fig.~\ref{fig:results}): we
get a very negative $\epsilon_{\rm f}^{-1}$ (panel~{\bf c}) not
accompanied by a large ${\cal A}_{\rm d}$ (panel~{\bf d}). The reason
is that this superlattice presents a small $\chi_{zz}$ (panel~{\bf
  b}), which yields large $\epsilon^{-1}$ and $|\epsilon_{\rm
  f}^{-1}|$. By the same token, having a globally soft superlattice
may result in a modest NC response of the f-layer, but this does not
necessarily imply a small VA. Hence, there is no reason to disregard
-- for VA purposes -- very responsive systems where small values of
$\epsilon^{-1}$ and $|\epsilon_{\rm f}^{-1}|$ have been measured or
computed\cite{zubko16,yadav19,das21}. Rather, we must focus on the
response difference between ferroelectric and dielectric layers, as
captured by the screening factor $\varphi_{\rm f}$.

Finally, let us stress that our conclusions for an idealized
superlattice apply to other materials too. First, note that an
infinite superlattice is equivalent to a ferroelectric/dielectric
bilayer contacted with good electrodes, so there is no net
depolarizing field. Second, NC is perfectly compatible with non-ideal
electrodes and depolarizing fields\cite{bratkovsky06}; in fact, these
fields reflect the electrostatic frustration at its origin. Hence, we
expect our conclusions to apply to real systems where the development
of an homogeneous polar state is precluded, including field-effect
transistors featuring a ferroelectric/semiconductor bilayer.

We hope this work will bring an impetus to the study of negative
capacitance, shifting the focus to the quantification and optimization
of voltage amplification.

\textbf{Methods}

The second-principles simulations are performed using the SCALE-UP
package~\cite{wojdel13,scaleup} and the same approach as previous
studies of PTO/STO superlattices\cite{das19,goncalves19,zubko16}. The
superlattice models are based on potentials for the pure bulk
compounds -- fitted to first-principles results\cite{wojdel13} -- and
adjusted for the superlattices as described in
Ref.~\onlinecite{zubko16}.

We study a collection of $n/m$ superlattices with layer thicknesses
$n, m=\{3, 6, 9\}$. Further, we consider an isotropic epitaxial strain
$\eta$ between $-1$~\% and 3~\%, where the STO square substrate (with
lattice constant of 3.901~\AA) is taken as the zero of strain.

We work with a simulation supercell that contains $8\times 8$
perovskite unit cells in the $xy$ plane (perpendicular to the stacking
direction). In the $z$ direction, only 1 superlattice period is
considered. Periodic boundary conditions are assumed.

In order to find the lowest-energy state of an $n/m$ superlattice at a
given $\eta$ and electric field value, we relax the atomic structure
by performing Monte Carlo (MC) simulated annealings. During the
annealings, all atomic positions and strains are allowed to vary,
except for the in-plane strains imposed by the substrate. From the
resulting atomic structures, we compute local electric dipoles within
a linear approximation (i.e., we consider the atomic displacements
with respect to the high-symmetry reference structure and multiply
them by their corresponding Born charge tensors), as customarily done
in second-principles studies\cite{zubko16,goncalves19}.

To compute responses, a small external field of 0.2~MV~cm$^{-1}$ is
considered. We checked that this field is small enough to obtain
susceptibilities and the other relevant quantities within a linear
approximation.

\vspace{5mm}{\bf\cabin Acknowledgements}

Work funded by the Luxembourg National Research Fund (FNR) through
grants INTER/RCUK/18/12601980 (M.G, J.\'I.) and
FNR/C18/MS/12705883/REFOX/Gonzalez (H.A.), and by the United Kingdom's
EPSRC through grant EP/S010769/1 (P.Z.).

\end{document}